\documentclass{svjour3}                     
\smartqed  
\usepackage{graphicx}
\begin{document}

\title{THE OBSERVATION OF EXOTIC RESONANCES FOR $\Lambda \gamma
$,$\Lambda\pi^{\pm}$,$\Lambda \pi^+\pi^-$, $\Lambda p$, $\Lambda pp$
and $p \gamma$  SPECTRA IN p + A COLLISION AT 10 GeV/c}
\thanks{Grants or other notes
}


\author{P.Zh.Aslanyan}


\institute{Joint Institute for Nuclear Research LHEP,  Dubna, Russia
\at
              Joliot-Curie 6, Moscow region, Russia\\
              Tel.: +7-49621-65757\\
              Fax: +7-49621-65180\\
              \email{paslanian@jinr.ru}         \\
 }

\date{Received: date / Accepted: date}

\maketitle

\begin{abstract}Strange multi-baryonic clusters are an  exiting  possibility to
explore the properties of cold dense baryonic matter. Recent results
on searches for exotic narrow resonances with $\Lambda$ hyperon
subsystems are reviewed. The observation of $\Sigma^0$,
$\Sigma^{*+}$(1385) and $K^{*\pm}$(892) well known resonances  from
PDG are a good tests of this method. The mean value of mass for
$\Sigma^{*-}(1385)$ resonance is shifted till mass of 1370 MeV/$c^2$
and width is two times larger than same value from PDG. Such of
behavior for width and mass of $\Sigma^{*-}(1385)$ resonance
interpreted as extensive contribution from stopping
$\Xi^-\to\Lambda\pi^-$  and medium effect with mass of
$\Sigma^{*-}(1385)$ resonance.In ($\Lambda\pi^+ \pi^-$) spectrum
there are observed enhancement signals from  $\Lambda ^*(1600)$,
$\Lambda ^*(1750)$ and $\Lambda ^*(1850)$ resonances. A number of
significant signals were found in the effective mass spectra of:
1)$\Lambda \gamma$,$p \gamma$ $\Lambda \pi^{\pm}$,$\Lambda \pi \pi$,
$\Lambda p$ and $\Lambda p p$ subsystems.

\end{abstract}

\section{Introduction}
\label{sec:1}

There are a few  actual problems of  nuclear and particle physics
which are concerning  a subject of study. These are
following\cite{knc}-\cite{pdg}: in-medium modification of hadrons,
the origin of hadron masses, the restoration of chiral symmetry, the
confinement of quarks in hadrons, the properties of cold dense
baryonic matter and non-perturbative QCD, the structure of neutron
stars.  Multi-quark states, glueballs and hybrids have been searched
for experimentally for a very long time, but none is established.

      Strange multibaryon states  with $\Lambda$- hyperon
       and $K_s^0$ –meson systems has been studied  by using
       data from 700000 stereo photographs or $10^6$ inelastic
        interactions which  was obtained from expose proton beams
        at 10 GeV/c to 2-m propane bubble chamber LHE,JINR\cite{v0} - \cite{hadron}. There are not
         sufficient experimental data concerning for strange-hyperons production
         in hadron -nucleus and nucleus-nucleus collisions over momentum range
         of  4-50 GeV/c. A survey for new experiments with much improved statistics
          compared  to those early data would hopefully resolve whether such
          "exotic" multi-quark hadron and baryon resonances exist.

 \section{($\Lambda, \pi^+$)  spectra}\label{sec:2} \indent

\begin{figure}
    \begin{center}
        {\includegraphics[width=55mm,height=40mm]{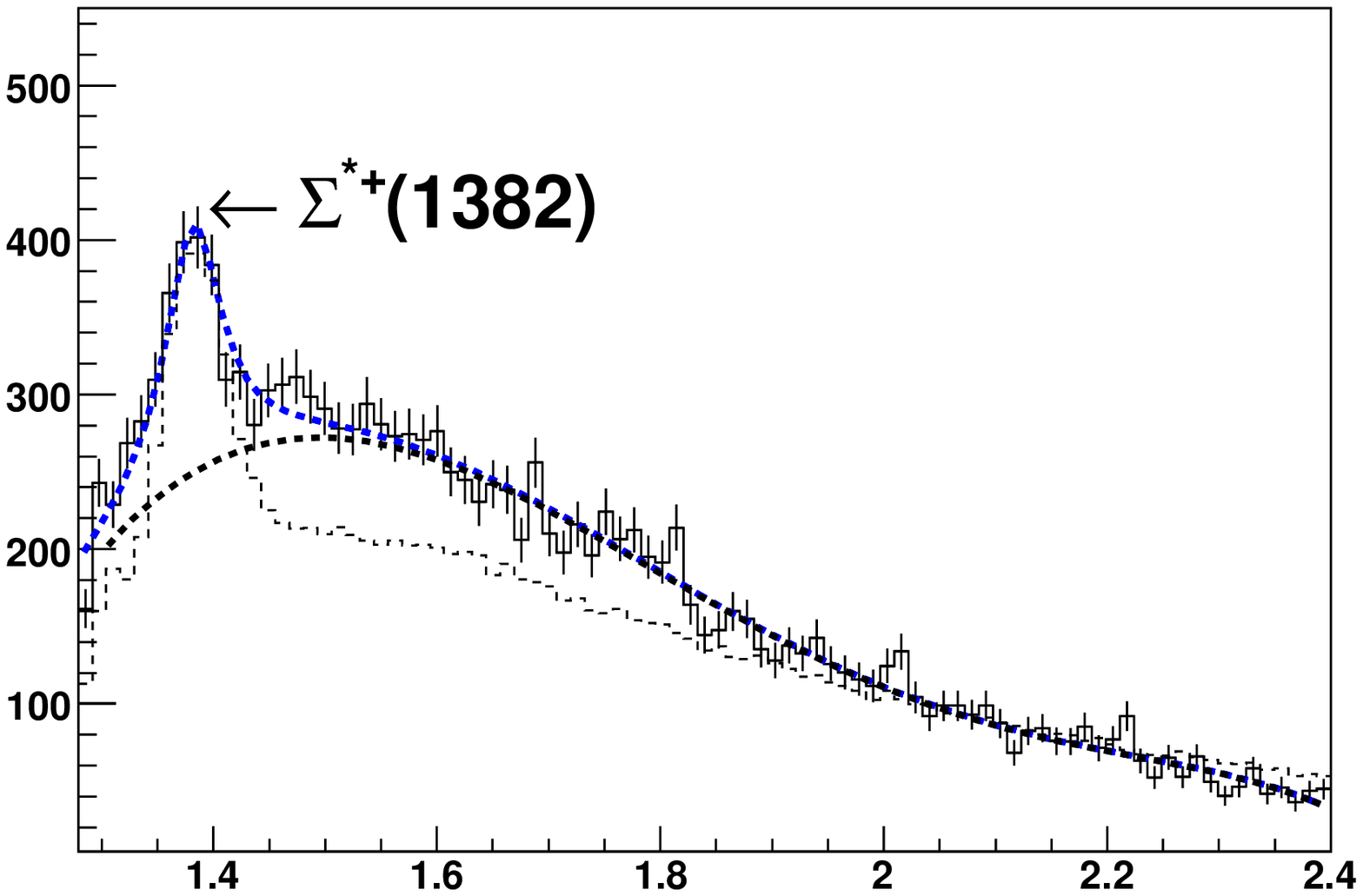}a)}
         {\includegraphics[width=55mm,height=40mm]{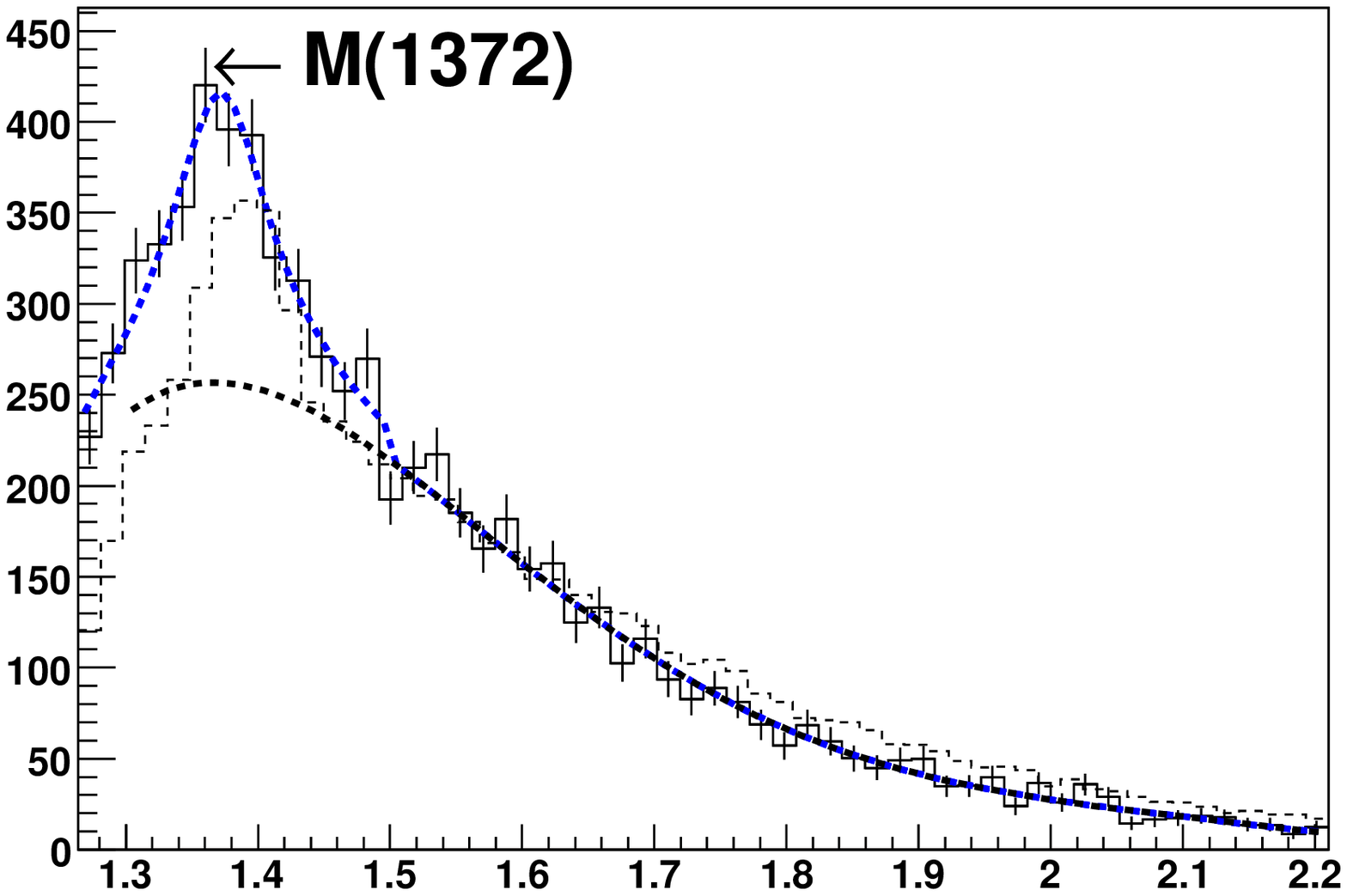}b)}

        \caption{\it a)The $\Lambda \pi^+$ - spectrum for all
        combinations with bin size of 12 MeV/$c^2$;
        b) $\Lambda\pi^-$  spectrum for all combinations with
  bin size of 14 MeV/$c^2$.
The simulated events by FRITIOF is the dashed histogram. The
background is the dashed curve.} \label{lpi}
  \end{center}
\end{figure}

The $\Lambda\pi^+$ effective  mass distribution for all 15444
combinations with bin size 12 MeV/$c^2$  has shown in Fig.\ref{lpi}a
\cite{hs07}-\cite{hadron}. The bin size is consistent with the
experimental resolution. The above dashed curve(Fig.\ref{lpi}a) is
the sum of the background  and 1 Breit-Wigner function
($\chi^2/N.D.F.=79/95$). The background (down dashed curve) is the 6
order polynomial function(Fig.\ref{lpi}a). Always the dashed
histogram  is the simulated background by the FRITIOF model in
figures. The mass resolution is equal to $\Delta M/M =0.7$\% in mass
range of $\Sigma^{*+}(1382)$. The decay width is equal to $\Gamma
\approx$ 45 MeV/$c^2$. The cross section of $\Sigma^{*+}(1382)$
production ($\approx$ 540 exp. events) is approximately equal to 0.9
mb for p+C interaction. This resonance with similar decay properties
for $\Sigma^{*+}(1382) \to\Lambda \pi^+$ identified which was a good
test of this method.

\section{ ($\Lambda, \pi^- $) spectra}\label{sec:3} \indent

The $\Lambda\pi^-$- effective  mass
distribution\cite{hs07},\cite{hadron} for all 6465 combinations with
bin sizes of 14 and 8 MeV/$c^2$ in Fig.\ref{lpi}b are shown. The
above dashed curve(Fig.\ref{lpi}b) is the sum of the background (by
the 8-order polynomial) and 1 BW function($\chi^2/N.D.F.=41/54$).
There is a significant enhancement in the mass range of 1372
MeV/$c^2$, with 12.5 S.D.,$\Gamma$ = 92 MeV/$c^2$. The cross section
of $\Sigma^{*-}$ production ($\approx$680 events) is equal to
$\approx$ 1.3 mb at 10 GeV/c for p+C interaction. The width for
$\Sigma^{*-}$ observed $\approx$2 times larger  than PDG value. One
of possible explanation is nuclear medium effects on invariant mass
spectra of hadrons decaying in nuclei\cite{sig}. There are also
significant enhancements in mass range of 1480 MeV/$c^2$, with 3.9
S.D..

\begin{figure}
    \begin{center}

          {\includegraphics[width=55mm,height=40mm]{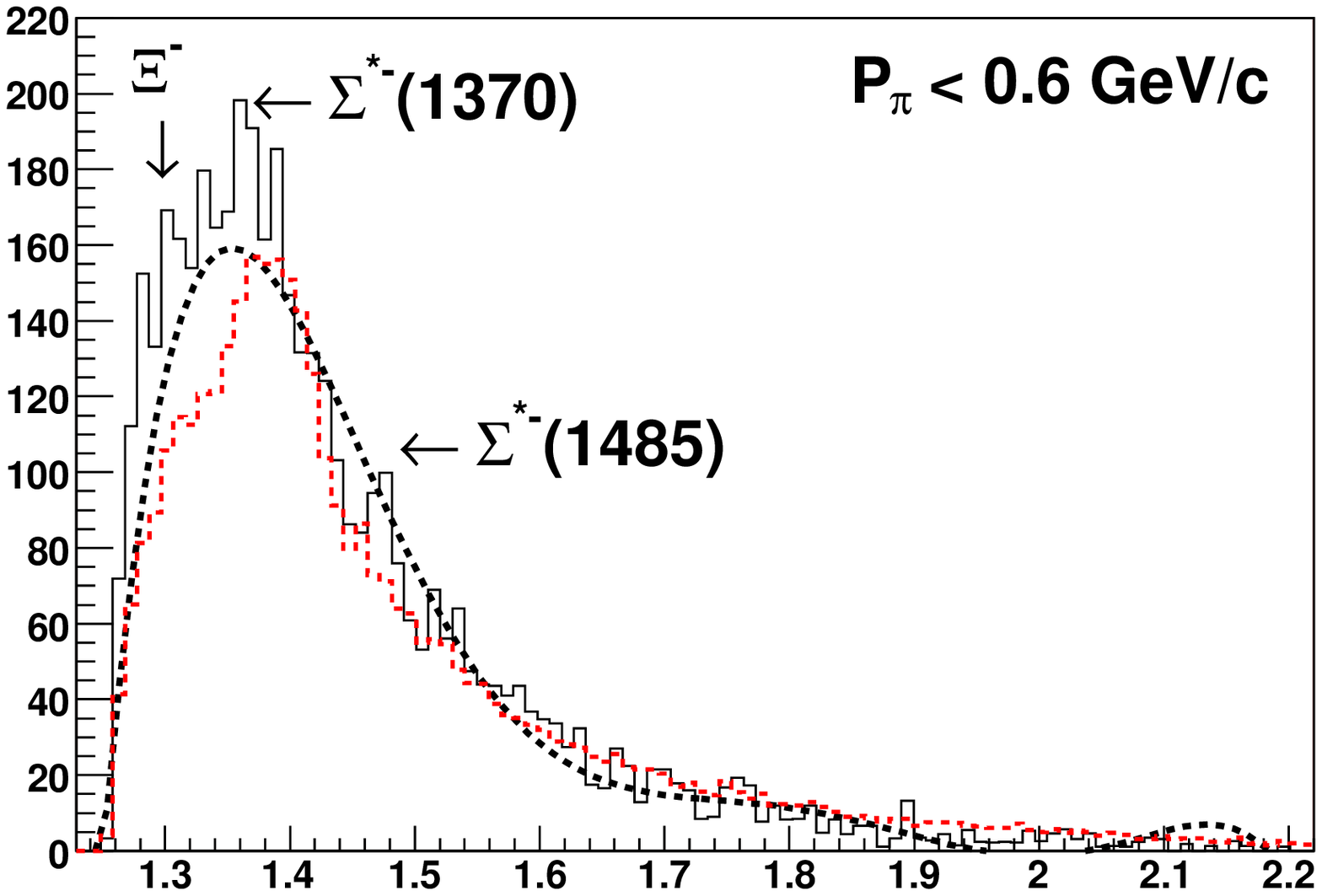}a)}
          {\includegraphics[width=55mm,height=40mm]{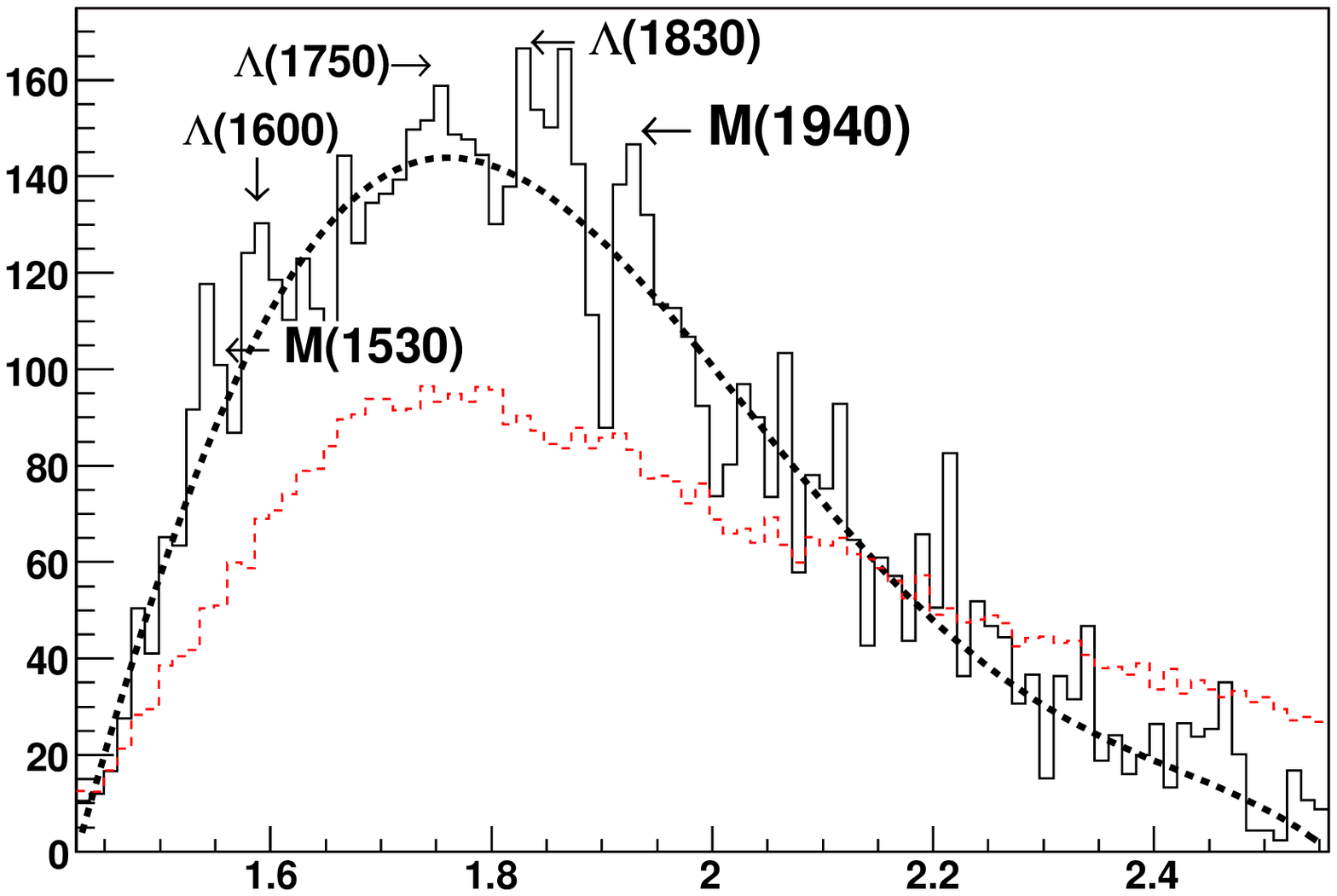}b)}
        \caption{\it a)$\Lambda \pi^-$ spectra in momentum range of $P_{\pi}<0.6 GeV/c$;
        b)The $\Lambda \pi^+ \pi^-$  spectra over momentum range of $P_{\pi^+}<1 GeV/c$
  with bin size 13 MeV/$c^2$;
 } \label{lpim}
  \end{center}
\end{figure}

Fig.~\ref{lpim}a shows the $\Lambda\pi^-$ - effective mass
distribution for 3829 combinations with bin size 10  MeV/$c^2$ over
momentum range of $P_{\pi}<$0.6 GeV/c, where removed combinations
from $K^-$. The dashed curve(Fig.\ref{lpim}a) is the background (6
-order polynomial). The observed width for $\Sigma^{*-}(1375)$ is
equal to $\approx$58 after fit of this effective mass distribution
by background+2BW function . In this case the width is  $\approx$1.5
times larger than PDG value.  There are small enhancements in mass
regions of 1320(4.1 S.D.) and 1480(4.2 S.D.)MeV/$c^2$.  With  bin
size 8 MeV/$c^2$ same distribution  is shown that there are small
enhancements in mass regions of: 1320 and 1485 MeV/$c^2$ too. In
this case, the peak in mass range of M(1372) is decayed into three
ranges:M(1320)+M(1372)+M(1385). Where M(1320)and M(1385) can
interpreted as contributions from $\Xi^-$ and $\Sigma^{*-}(1385)$,
respectively. Then M(1372) peak can interpreted as contribution from
phase space or medium effect with $\Sigma^{*-}(1385)$ in carbon
nucleus. The number of $\Xi^-$ stopped in nuclear medium is equal to
$\approx$ 60 events for p+propane interaction.

The total  number of events  by weak decay channel with $\Xi^-$ is
equal to 16 (w=1/$e_{\Lambda}$ =5.3, where is a full geometrical
weight of registered for $\Lambda$s)\cite{H1}. Then experimental
cross section for identified $\Xi^-$ by weak decay channel\cite{H1}
is more than two times larger than from calculation by FRITIOF. The
observed total experimental cross section for $\Xi^-$  is more than
4 times larger than the cross section which is obtained by FRITIOF
model for same experimental conditions. Figures ~\ref{lpim}a shows
that there is the significant enhancement in mass ranges of
$\Sigma^{*-}$(1485)\cite{pdg} which is agreed with the reports from
SVD2 and COSY collaborations but none established.

\subsection{($\Lambda,\pi^+,\pi^- $) spectrum}\label{sec:4}
 \indent

The $\Lambda\pi^+\pi^-$  effective  mass
distribution\cite{hs07},\cite{hadron} for all 6483 combinations with
bin size 13  MeV/$c^2$ has shown in Figure \ref{lpim}b. The dashed
curve and dashed histogram are backgrounds by the polynomial
function and FRITIOF model, respectively.   There are enhancements
in mass regions of 1530,1600, 1750, 1830-1860,(1930-1940), 2030 and
2250 MeV/$c^2$ which can interpreted as a reflection from resonances
of $\Lambda^*$(1520),$\Lambda^*$(1600),$\Lambda^*$(1750)
$\Lambda^*$(1850)and $\Xi^*(1830)(\to \Lambda\overline{K^0_s}$),
$\Sigma^*$(1940)(and $\Xi^*(1950)\to \Lambda\overline{K^0_s}$),
$\Sigma^*$(2030)(and $\Xi^*(1950)\to \Lambda\overline{K^0_s}$) and
$\Sigma^*$(2250) from PDG.

 \section{($\Lambda, p$) and ($\Lambda, p, p $) spectra}\label{sec:5} \indent

\begin{figure}
    \begin{center}

 {\includegraphics[width=60mm,height=40mm]{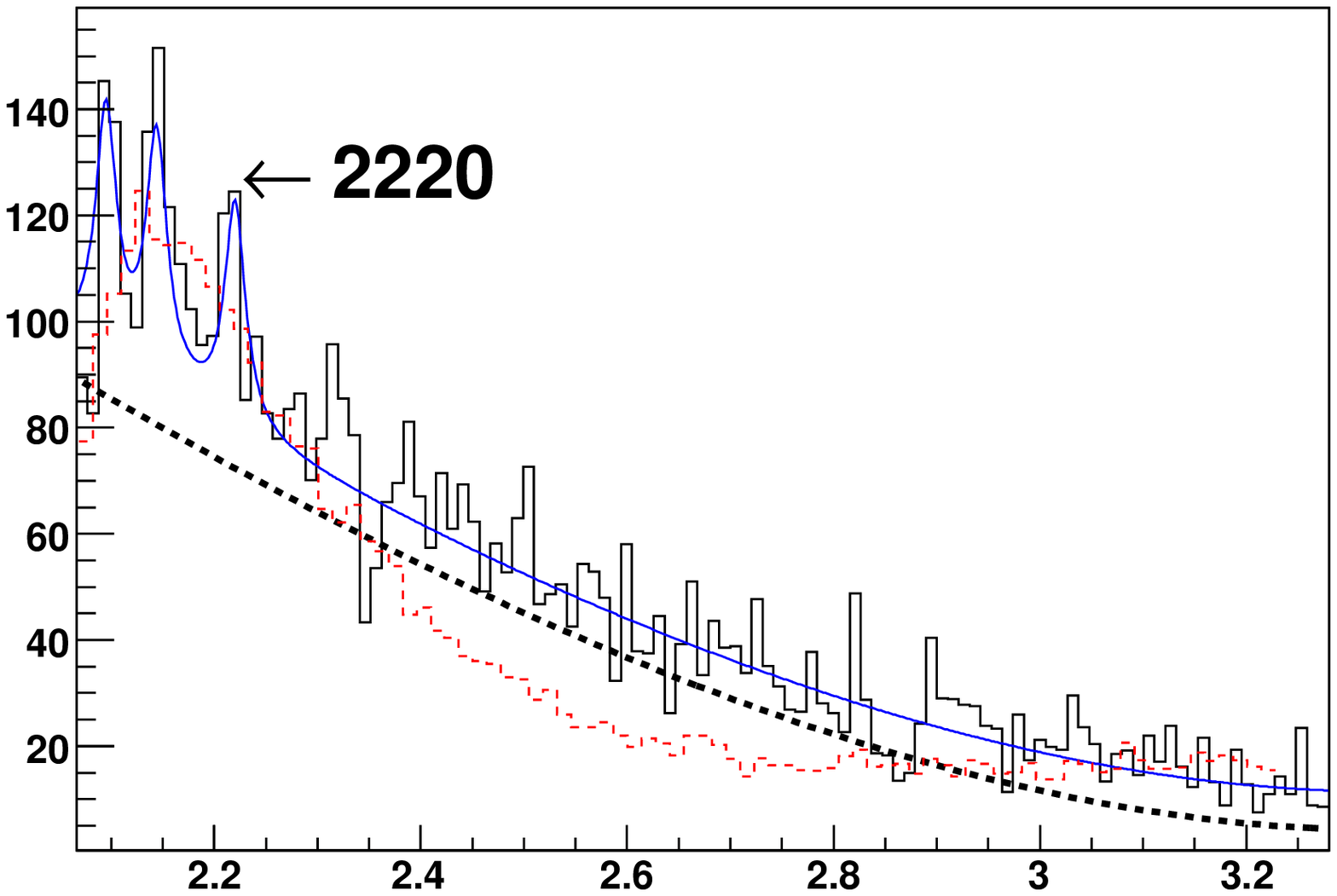}a)}
          {\includegraphics[width=55mm,height=40mm]{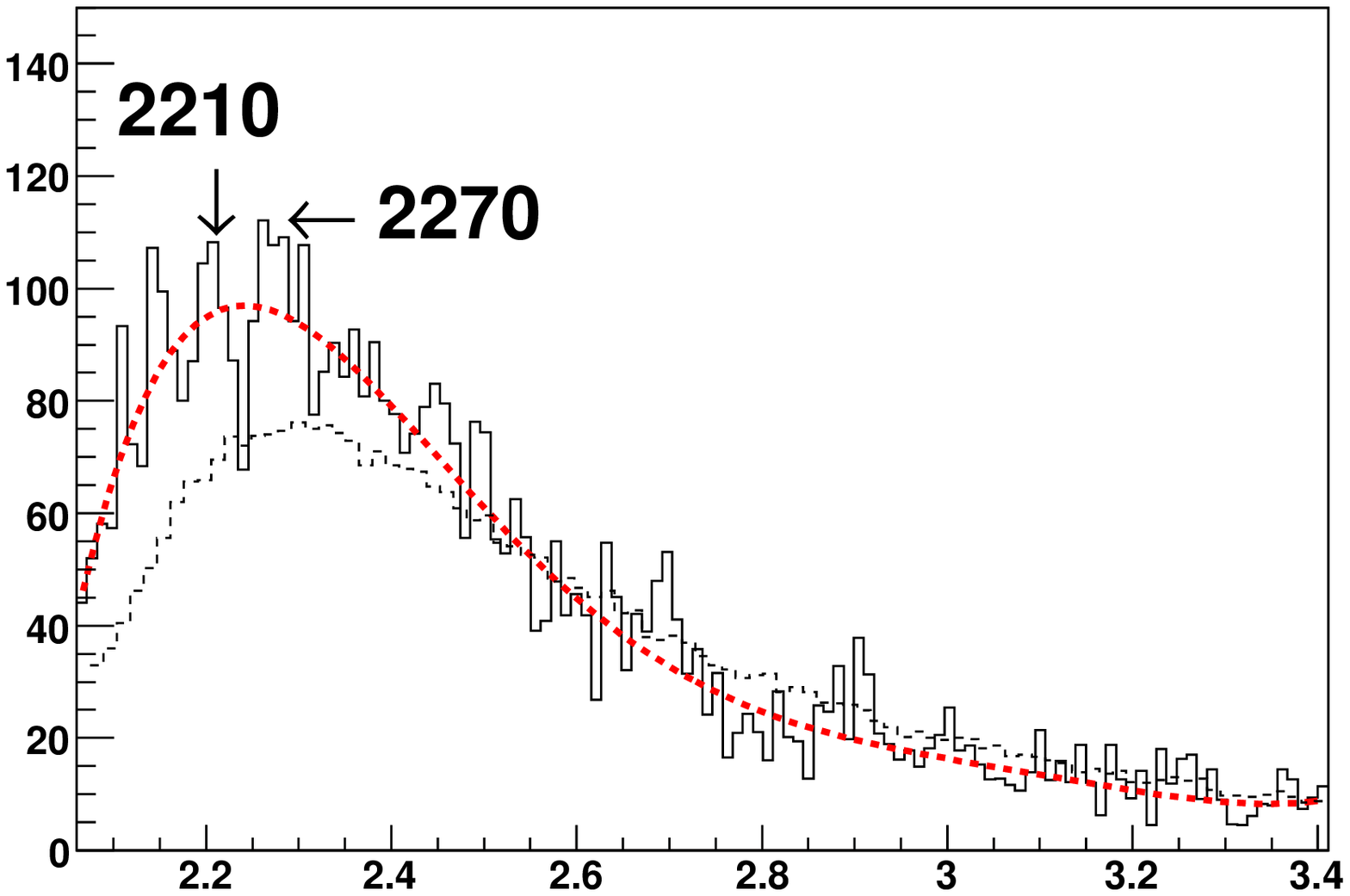}b)}
        \caption{\it a) the $\Lambda p$ spectrum  with stopped protons in momentum range of 0.14$<P_p<$0.30 GeV/c;
        b)the $\Lambda p$  spectrum  for relativistic positive tracks in range of $P_p>$1.5 GeV/c
 } \label{lp}
  \end{center}
\end{figure}

Figure \ref{lp}a  shows the invariant mass of 4011($\Lambda
p$)combinations with bin size 15 MeV/$c^2 $ for stopped protons in
momentum range of 0.14$< P_p<$ 0.30 GeV/c
\cite{pomer}-\cite{hadron}. The dashed curve is the sum of the
8-order polynomial  and 4 Breit-Wigner curves with $\chi^2=30/25$
from fits. The significant peak at
 mass range of 2220 MeV/$c^2$ (6.1 S.D.), $B_K$ ~ 120 MeV was specially
stressed by Professor T. Yamazaki on $\mu$CF2007, Dubna,
June-19-2007 that is conform with KNC model\cite{knc} prediction by
channel of $K^- pp \to \Lambda $p .

 The $\Lambda p$ effective mass distribution for
4523 combinations with relativistic protons over a momentum of P
$>$1.5 GeV/c is shown in Figure \ref{lp}b, where removed undivided
($\Lambda K^0_s$). The solid curve is the 6-order polynomial
function($\chi^2$/n.d.f=271/126). The background for analysis of the
experimental data are based on FRITIOF and the polynomial method.
There are significant enhancements in mass regions of 2150(4.4
S.D.), 2210(3.8 S.D.), 2270(3.4 S.D.),2670 (3.1 S.D.)and 2900(3.1
S.D.)MeV/c$^2$. The observed peaks for combinations with
relativistic protons P $>$1.5 GeV/c agree with peaks for combination
with identified protons and with stopped protons.

\begin{figure}
\begin{center}
          {\includegraphics[width=30mm,height=50mm]{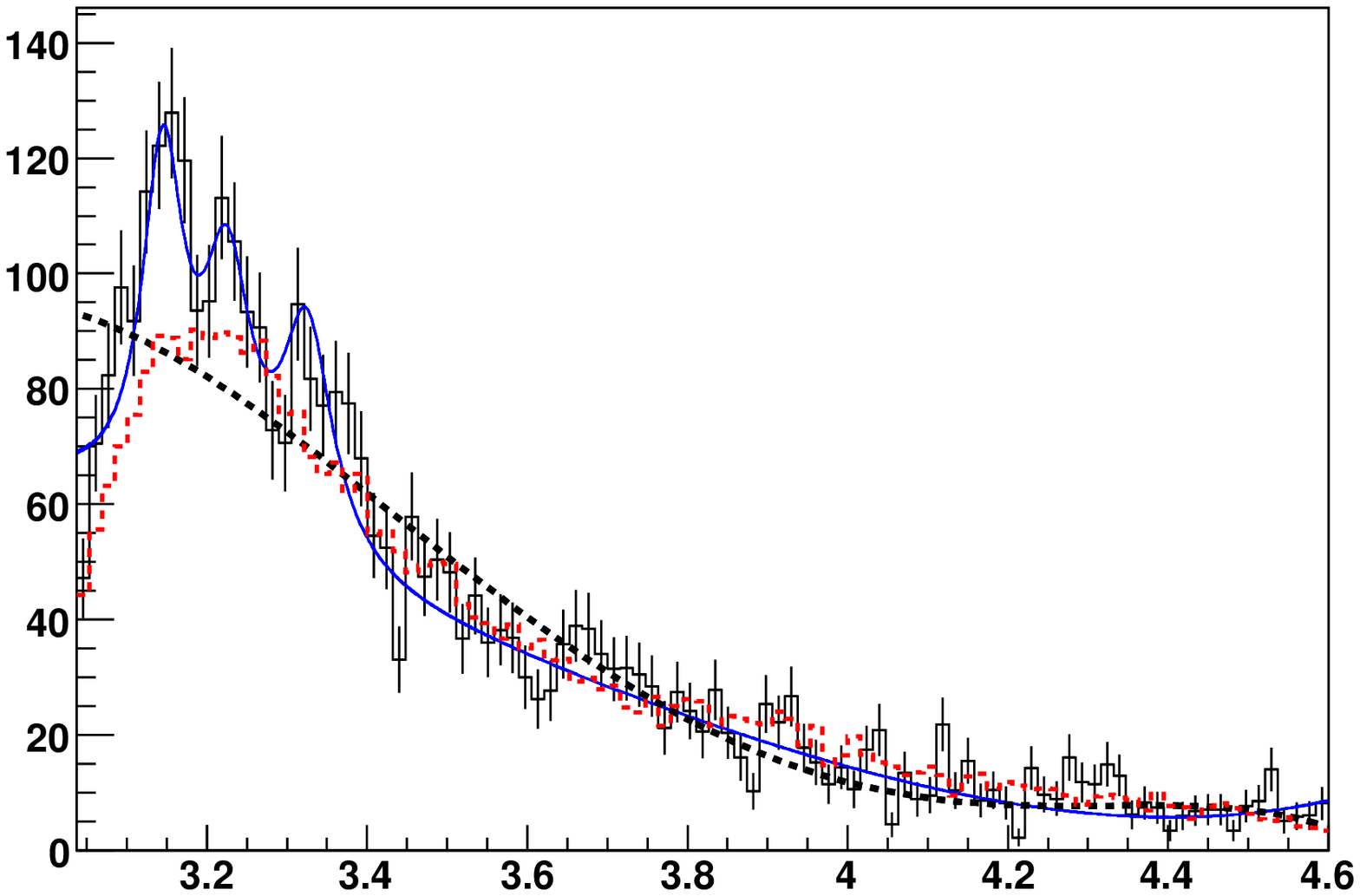}a)}
          {\includegraphics[width=30mm,height=50mm]{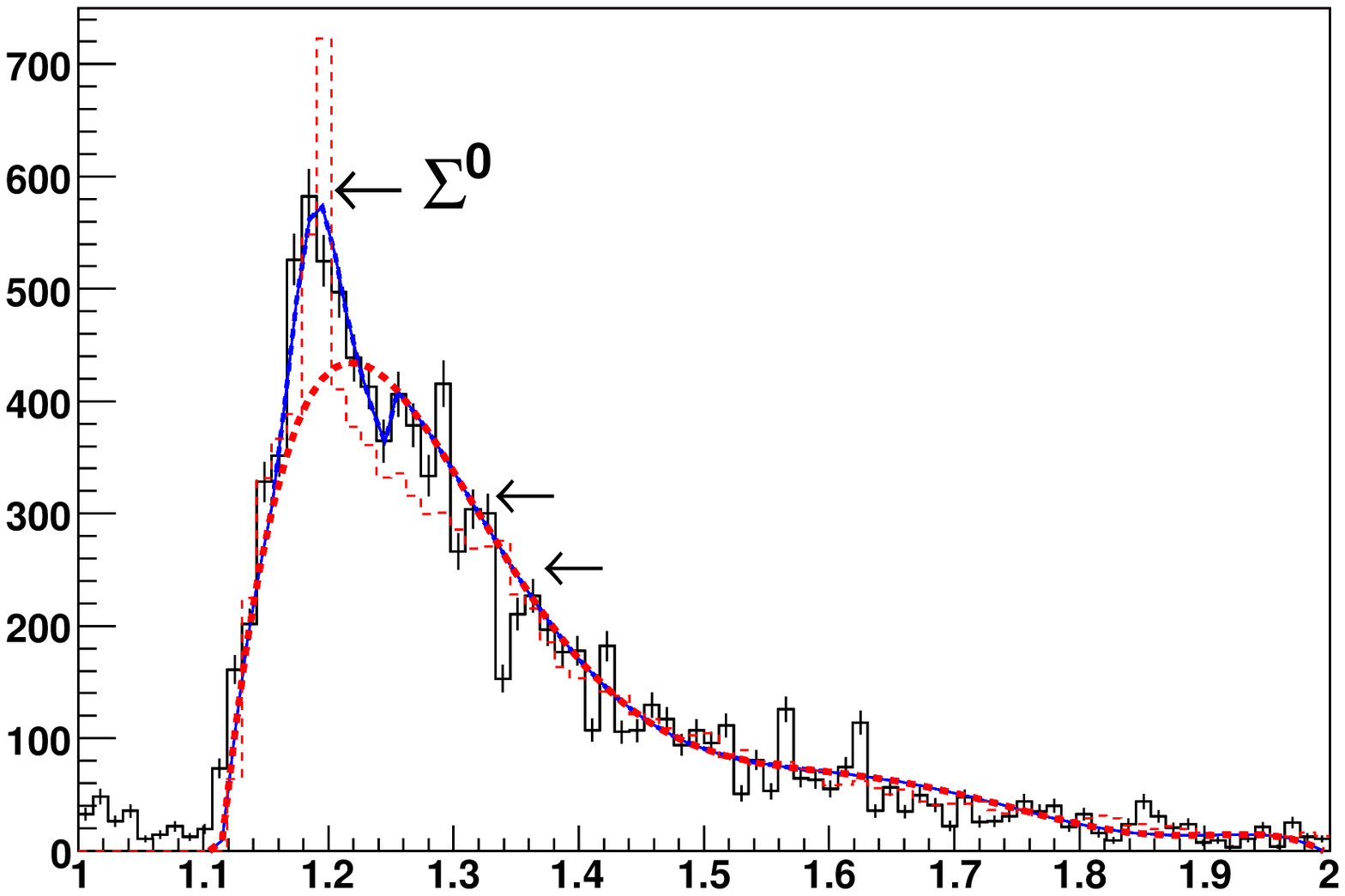}b)}
          {\includegraphics[width=50mm,height=50mm]{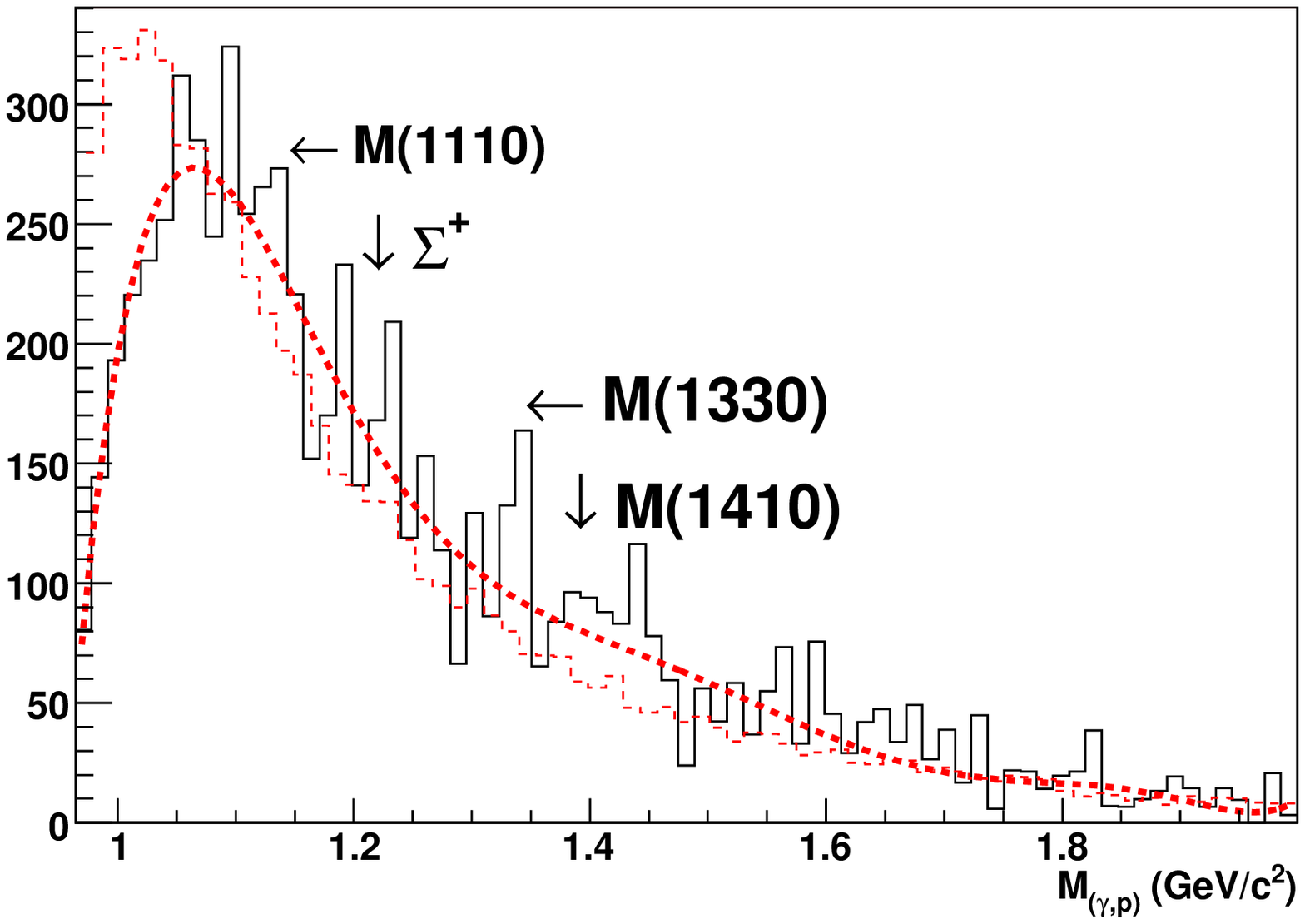}c)}
   \caption{\it a)The $\Lambda p p$ spectrum for identified
  protons $P_p<$0.9 GeV/c;b) the $\Lambda \gamma$ spectrum for 2630 combination
  with bin size of 12 MeV/$c^2$; c)the $\gamma p$ spectrum
  for 1727 combination with protons at momentum range of $P_p<$0.9 GeV/c.   }
 \label{lg}
 \end{center}
 \end{figure}

The $\Lambda pp$   effective  mass distribution for 3401
combinations for identified protons with a momentum of  $P_p <$0.9
GeV/c is shown in Figure \ref{lg}a)\cite{hs07}-\cite{hadron}. The
dashed curve is the 6-order polynomial function ($\chi^2$/n.d.f
=245/58, Fig.\ref{lg}a ).In this case the analysis of the
experimental data are based on backgrounds by FRITIOF and the
polynomial method. There is significant enhancements in mass regions
of 3145 MeV/$c^2$(6.1 S.D.) and with width 40 MeV/$c^2$. There are
small enhancements in mass regions of 3225(3.3 S.D.), 3325(5.1
S.D.), 3440(3.9 S.D) and 3652MeV/$c^2$(2.6 S.D.). These peaks from
$\Lambda p$ and $\Lambda p p$ spectra were partly conformed with
experimental results from   FOPI(GSI),
FINUDA(INFN), OBELIX(CERN) and E471(KEK).\\

\section{($\Lambda\gamma $) spectrum}\label{sec:6}
 \indent

The $\Lambda \gamma$ (preliminary)  effective  mass distribution for
2630 combinations  is shown in Figure \ref{lg}b)with total
geometrical weights for $\Lambda$ and $\gamma$. The cross section of
$\Sigma^0$ production ($\approx$720 events, with geometric weights,
$<w_{\gamma}>$=4.1) is equal to 1.3 mb at 10 GeV/c for p+C
interaction at 10 GeV/c  which is more 1.5 times larger than
simulated cross section by FRITIOF.  The observed width of
$\Sigma^0$ is $\approx$ 2 times larger than  same value from
simulation by FRITIOF. There are enhancements in mass range of
1290,1321 and $\Sigma^{*0}$(1385) at bin size 12 and 9 MeV/$c^2$
what are interpreted as reflection from enhancement productions for
well known $\Xi^0$ and $\Sigma^{*0}$(1385) hyperons in effective
mass spectrum. There are same enhancements in the $\Lambda\gamma$
effective mass distribution without geometric efficiency for
$\Lambda$  and $\gamma$.

\section{( $\gamma p $) spectrum}\label{sec:7}
 \indent

The $\gamma p$ (preliminary)  effective  mass distribution for 1727
combinations at momentum of $p_p<0.9 GeV/c$ in reaction p+A$\to
K^0_s \gamma$pX with bin size 12MeV/$c^2$ is shown in Figure
\ref{lg}c)with geometric weight for $\gamma$ ($<w_{\gamma}>
\approx4.1$). The observed peaks in mass range of $\Sigma^+$(1189)(6
S.D.) and 1230 MeV/$c^2$ is reflection from decay p$\pi^0$(the
probability for this mode is 51.57 \%). The peak in mass range of
1230 MeV/$c^2$ can interpreted as shift mass of $\Sigma^+$ in versus
of registration mode by $p\gamma$. The significant signals have
observed in mass range of 1330 MeV/$c^2$ (7 S.D.) There are small
enhancements in mass range of 1050, 1110 and 1410 MeV/$c^2$.A
$\Lambda$ state at 1330 MeV/$c^2$ have been suggested in
\cite{azimov}. This distribution have same behavior without
geometric weights for $\gamma$.  There are same enhancements in the
effective mass distribution for $\gamma p$(3308 comb.) without cut
of over $P_p<0.9$ GeV/c momentum range.

\section{Conclusion} \label{sec:7}

$\bullet{}$ The observation of $\Sigma^0$,$\Sigma^{*+}$(1385) and
$K^{*+}$(890)states are a good tests for applied method.

$\bullet{}$The mass  of the exited $\Sigma^{*-}$(1385) is shifted in
mass range of 1372 MeV/c$^2$  and  the width is  two time
larger than the  PDG (preliminary result).\\

$\bullet{}$The production of stopped in medium $\Xi^-\to \Lambda
\pi^-$ is more than 4 times larger than expected geometrical cross
section for p+propane interaction (preliminary result).\\

$\bullet{}$ Peaks   for ( $\Lambda,p$) and ($\Lambda$,p,p) spectra
\cite{pomer}-\cite{hadron} are agreed with experimental data from
the recently reports of FOPI, KEK, OBELIX, FINUDA collaborations,
but there are some conflicting with peak positions or widths.

$\bullet{}$ There are signals in mass range of $\Sigma^*$(1480) and
M(1330)  by channels of  ($\Lambda,\pi$), ($\Lambda,\gamma$) and (p
,$\gamma$)(preliminary), respectively.

$\bullet{}$ There are enhancement production for all observed
hyperons and  by channel ($\Lambda\pi^+\pi^-$) too.

$\bullet{}$The search and study for exotic  strange multiquark
states with  $\Lambda$ and  $K_s^0$ subsystems at
  FAIR(GSI), JPARC(KEK), FINUDA(INFN)  and  MPD( NICA,JINR)
   can get a valuable information about their nature, properties
   and it will be a test for observed data on PBC.
   Higher statistics for experiments with mass resolution ($\approx$1\%) are needed.

\begin{acknowledgements}\label{9}
 My thanks EXA/LEAP'08 Organizing Committee and personally E. Widmann for providing the excellent, warm and stimulat-
ing atmosphere during the Conference and for the financial support.
 \end{acknowledgements}

%





\end{document}